\documentclass{aps2010}

\usepackage{graphicx}
\usepackage{rotating}

\title{Advanced Multi-beam Spectrometer for the Green Bank Telescope}

\author[org1]{\emph{\underline{D. Anish Roshi}}}
\author[org1]{\emph{Marty Bloss$^{1}$, Patrick Brandt}}
\author[org3]{\emph{Srikanth Bussa$^{1,}$}}
\author[org4]{\emph{Hong Chen}}
\author[org2]{\emph{Paul Demorest$^2$, Gregory Desvignes$^4$, Terry Filiba$^4$, Richard J. Fisher$^2$,
John Ford$^1$, David Frayer$^1$, Robert Garwood$^{2}$, Suraj Gowda$^4$, Glenn Jones$^{1,5}$, Billy Mallard$^4$, 
Joseph Masters}}
\author[org1]{\emph{Randy McCullough$^1$, Guifre Molera$^4$, Karen O'Neil$^1$, Jason Ray$^1$, Simon Scott$^4$,
Amy Shelton$^1$, Andrew Siemion$^4$, Mark Wagner$^4$, Galen Watts$^1$, Dan Werthimer$^4$, Mark Whitehead}}

\address[org1]{National Radio Astronomy Observatory (NRAO), P. O. Box 2, Green Bank, 
West Virginia 24944, aroshi@nrao.edu}
\address[org2]{National Radio Astronomy Observatory (NRAO), Charlottesville, VA 22903-2475}
\address[org3]{University of Akron, Akron, Ohio 44325}
\address[org4]{University of California, Berkeley, CA 94720}
\address[org5]{Caltech, Pasadena, CA 91125}

\begin{document}%
\maketitleblock  
\begin{abstract}
A new spectrometer for the Green Bank Telescope (GBT) is being built jointly by the NRAO and
the CASPER, University of California, Berkeley.
The spectrometer uses 8 bit ADCs and will be capable of processing up to 1.25 GHz bandwidth from 8
dual polarized beams. This mode will be used to process data from focal plane arrays.
The spectrometer supports observing mode with 8 tunable digital sub-bands within the 1.25 GHz bandwidth. 
The spectrometer can also be configured to process a bandwidth of up to 10 GHz with
64 tunable sub-bands from a dual polarized beam.
The vastly enhanced backend capabilities will support several new science projects
with the GBT. 
\end{abstract}

\section{Introduction}

The Robert C. Byrd Green Bank Telescope (GBT) is the premiere single-dish radio
telescope operating at meter, centimeter and millimeter wavelengths [1]. 
With an off-axis parabolic reflector of size 100 m $\times$ 110 m, the GBT
is the largest fully-steerable radio telescope in the world. Its unblocked 
aperture gives it the sensitivity of a much larger antenna, flat spectral baselines and 
higher sensitivity to low surface brightness
emission. The GBT is located at the National Radio Astronomy Observatory (NRAO)\footnote{National Radio 
Astronomy Observatory is a facility of the National Science Foundation operated under 
cooperative agreement by Associated Universities, Inc.} site 
in Green Bank, West Virginia, USA. It is being used for a variety of unique experiments
over a broad range of science. 

Since its commissioning in 2000, the GBT's capabilities have been greatly expanded
in frequency coverage as well as the instantaneous field of view for observations. 
The field of view is enhanced using focal plane arrays. The new 4mm receiver
is an example of expansion in frequency coverage of the GBT. So far observers have been 
using all these receivers with an existing spectrometer backend, which  
has only 3 level sampling, 800 MHz bandwidth and a minimum integration 
time of one sec. The bandwidth supported by this spectrometer for focal plane arrays 
is $\le$ 50 MHz. Thus the capabilities of the existing spectrometer do not match 
with those of the new receivers. It is now time to upgrade the backend capability 
of the GBT. 

The National Science Foundation Advanced Technologies and Instrumentation (NSF-ATI)
program has funded a new spectrometer backend for the GBT. This spectrometer is 
being built by the CICADA collaboration -- a collaboration between the 
NRAO and the Center for Astronomy Signal Processing 
and Electronics Research (CASPER) at the University of California, Berkeley.
The new spectrometer vastly enhance the capability of the GBT to support
several unique science projects. The projects include mapping 
temperature and density structure of molecular clouds;
searches for organic molecules in the interstellar medium; determination of the
fundamental constants of our evolving Universe; redshifted spectral features from galaxies
across cosmic time and survey for pulsars in
the extreme gravitational environment of the Galactic Center.

\section{GBT spectrometer design}

\begin{figure}[t]
\includegraphics[width=4.0in, height=7.0in, angle=-90]{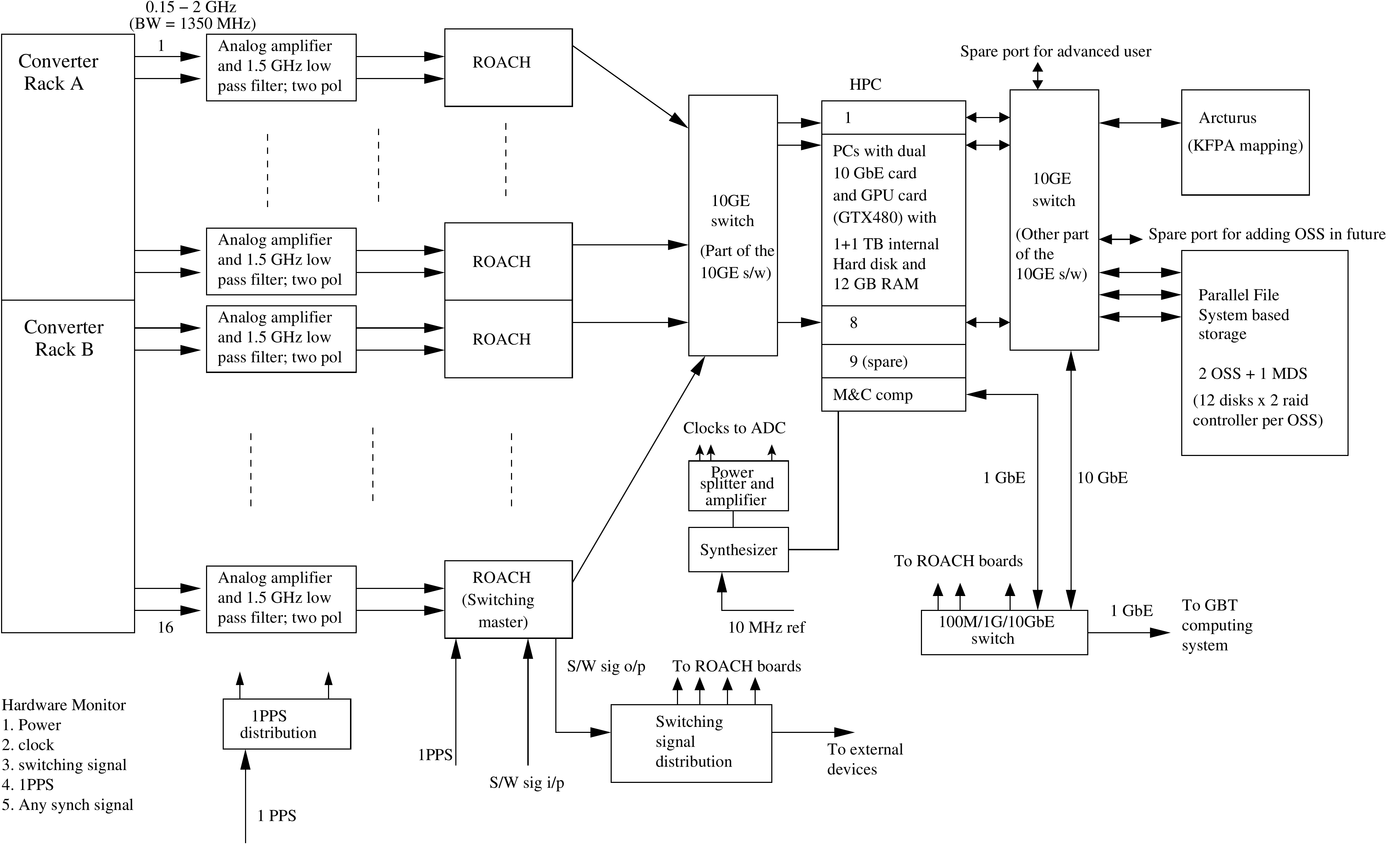}
\caption{A block diagram of the GBT spectrometer}
\label{fig1}
\end{figure}

One of the design considerations for the new spectrometer is to process larger bandwidth (1.25 GHz)
with the focal plane array (FPA) receivers. The K-band (18 -- 26.5 GHz) FPA 
of the GBT has 7 feeds. Thus the spectrometer is designed to 
process 8 (7 + 1 spare) dual polarized signals. These signals are processed 
by 8 independent spectrometer units working in parallel.  
A block diagram of the spectrometer is shown in Fig.~\ref{fig1}. 
The 8 spectrometer units are connected
to the Converter Racks (CR) outputs, which is the last stage of the GBT's intermediate
frequency (IF) system. The spectrometer can broadly
be divided into two parts: (1) FPGA based hardware and (2) pipeline
computing facility. CASPER's ROACH\footnote{see http://casper.berkeley.edu/wiki/ROACH} 
board is used as the FPGA based
hardware. The pipeline computing will be implemented using a cluster
computing facility (HPC).

The analog signals from the CR, after amplification and band limiting  
to 1.5 GHz, are fed to  8-bit ADCs attached to the ROACH boards. 
The sampling clock for the ADC is generated using a synthesizer, which is locked to
the observatory 10 MHz standard. The
FPGAs in the ROACH board process the digitized signal and send spectra and
cross correlations of the two polarizations to the HPC
through a 10 GbE switch. The HPC consists of 8 (+ 1 spare) independent PCs.
Each PC will have a dual port 10 GbE card and a GPU GTX480 card. The GPU will be used for
implementing observing modes with bandwidth $\le$ 250 MHz and
the 8 sub-band capability (see Section~\ref{mode}). One of the ports of the 10GbE card
will receive data from the ROACH boards and the second port will send
data to the data storage system. The HPC is connected to the data storage
system through the same 10 GbE switch to which the ROACH boards are connected.
The data storage system will be a parallel file system (eg. lustre) based
storage and will support $\sim$ 100 TB of disk space. For all the
specified observing modes data will be written to this storage system.
The GBT users will be accessing data from this storage system through
a 1 GbE link as shown in Fig~\ref{fig1}.

The monitor and control (M\&C) of the whole spectrometer is done through a dedicated
computer. The M\&C system will obtain data from the ROACH board through
a 1 GbE (or 100 MbE) network. A 10 GbE link to this network provides
the M\&C information from the HPC.

A 1 PPS signal from the observatory time standard is connected to the 
ROACH boards. This 1 PPS is the main synchronization
signal for the spectrometer. Switching signals are used during observations
to synchronize spectral measurements with noise and/or frequency
switching. The internal switching signals are
generated by one of the ROACH boards (Switching Master) and will be
sent to other ROACH boards through a distribution system. The
Switching Master will also accepts external switching signals
and will send them to other boards through the same distribution system.

\begin{figure}[t]
\includegraphics[width=3.0in, height=7.0in, angle=-90]{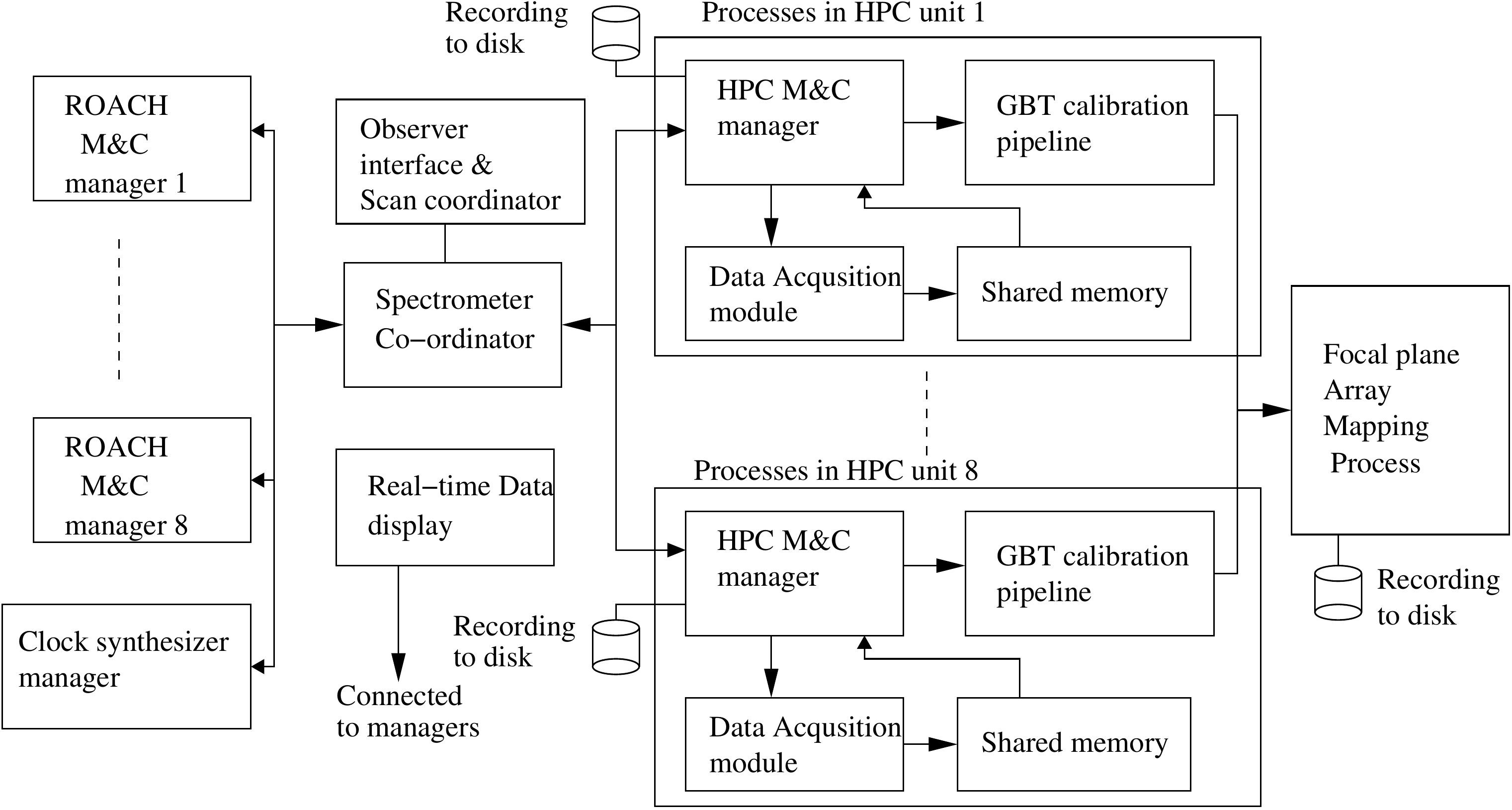}
\caption{Software architecture of the GBT spectrometer}
\label{fig2}
\end{figure}

The spectrometer software architecture is shown in Fig.~\ref{fig2}.
The software components can be broadly divided into two parts -- the 
M\&C and the observing pipeline. The M\&C
is designed based on the existing GBT software system.
In this system each hardware device is controlled by a `manager'. The
manager also sends the monitoring information at regular intervals. 
They also have the functionality to record data to a FITS file.
In the spectrometer design, the ROACH M\&C managers control the ROACH boards
and the HPC M\&C managers control the data acquisition and flow in the HPC units. 
The spectra from the ROACH are averaged to the required integration time and put in a 
shared memory by the data acquisition module. The HPC M\&C manager reads the data from the 
shared memory and records to the disk storage system. 

For FPA observations, the HPC M\&C manager sends data to the
data processing pipeline. This pipeline will produce a uniformly gridded, calibrated 
spectral map of the observing region as its end product. The operations of this pipeline
can be divided into beam based calibration and mapping process.  
The beam based calibration process (GBT calibration pipeline) will be
implemented in the HPC. After calibration, the HPC sends the data to a
computer (Arcturus) where the spectral image is made by the mapping process. 

\section{Observing Modes}
\label{mode}

\begin{figure}[t]
\includegraphics[width=3.0in, height=7.0in, angle=-90]{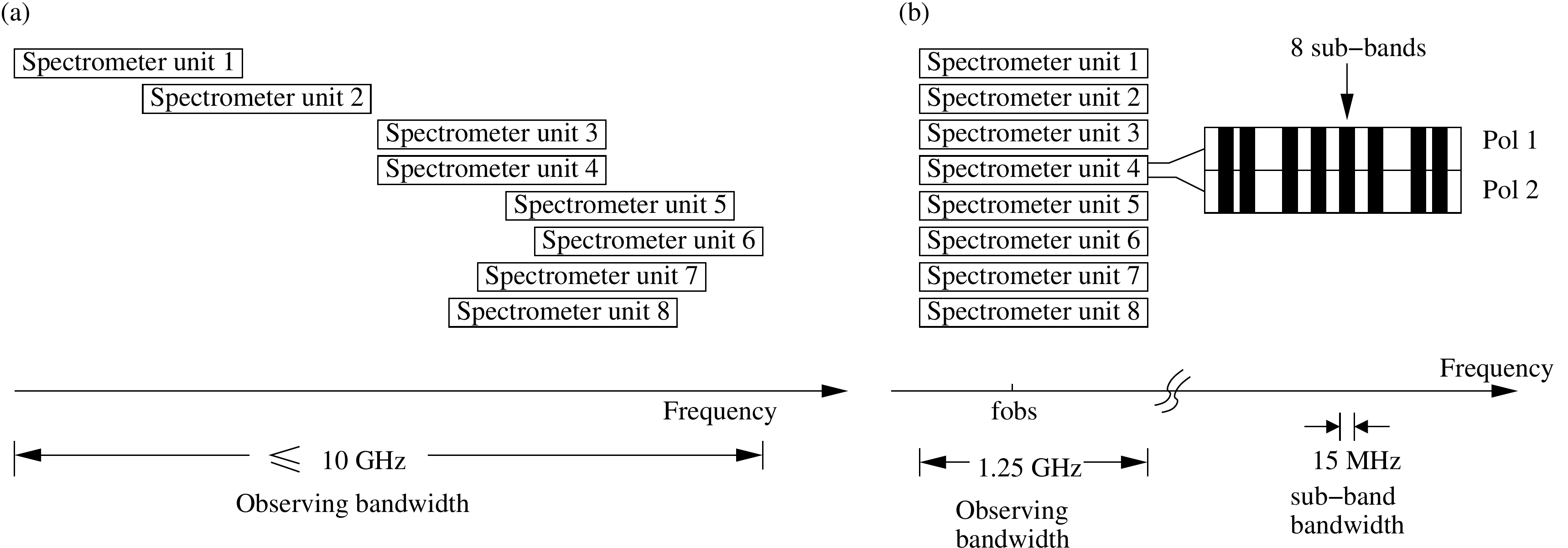}
\caption{(a) An example of the spectrometer configuration 
for single beam, dual polarized observations. The 8 spectrometer units
can be tuned to different frequencies
covering a bandwidth of up to 10 GHz. Each unit processes signals from 
two polarizations.  
(b)An example of the spectrometer mode for focal plane array observations.
Each spectrometer unit processes a bandwidth of 1.25 GHz centered
at the same observing frequency.
In this example, there are 8 sub-bands, each of 15 MHz bandwidth, 
tuned to different frequencies within the 1.25 GHz bandwidth. 
}
\label{fig3}
\end{figure}

The spectrometer supports a variety of observing modes. The major
modes of operation are single and 8 sub-band modes. In
single sub-band mode the maximum bandwidth a spectrometer unit can 
process is 1.25 GHz. There will be 1024 spectral channels across this
bandwidth which gives a spectral resolution of $\sim$ 1.5 KHz. The
minimum integration time supported by this mode is 0.5 msec. The
minimum bandwidth provided in single sub-band mode is 1 MHz
with spectral resolution of 30 Hz (ie 32768 channels). The
minimum integration time for this mode is about 10 msec. 
The spectrometer will support 8 fully tunable sub-bands 
within the 1.25 GHz bandwidth (see Fig.~\ref{fig3}b). The 
center frequency of the sub-band is tunable to an accuracy of 10 KHz.
The sub-band bandwidth varies from 30 MHz to 1 MHz. The number
of spectral channels per sub-band for these different bandwidths is 4096. 

The 8 spectrometer units can be configured in different 
observing modes. They can also be used together to process  
a contiguous bandwidth of up to 10 GHz. 
An example of using the spectrometer in this 
configuration is shown in Fig.~\ref{fig3}a. 
The total number of sub-bands available when all the 8 
units are used together is 64. 
For FPA observations, all the 8 spectrometer units will usually 
be configured to the same observing mode (see Fig.~\ref{fig3}b).

\section{Acknowledgments}
We thank Chris Clark, Eric Sessoms, Jay Lockman, Mark Clark, Richard Lacasse, Roger Norrod,
Ron Maddalena, Steven White and Wolfgang Baudler for their informative comments and suggestions 
at various stages of the spectrometer project. 

\section{References}
1. Prestage, R. M., et. al.,``The Green Bank Telescope," \emph{Proceedings of
the IEEE}, Vol 97, Issue 8, 2009, pp. 1382-1390 \\

\end{document}